\documentstyle[twocolumn,aps]{revtex}

\begin{document}
\draft
\title{Entanglement measures with asymptotic weak-monotonicity as lower
(upper) bound for the entanglement of cost
(distillation)} 
\author{Won-Young Hwang \cite{email}, and
Keiji Matsumoto \cite{byline}}
\address{IMAI Quantum Computation and Information Project,
 ERATO, Japan Science Technology Cooperation,
Daini Hongo White Bldg. 201, 5-28-3, Hongo, Bunkyo-ku,
Tokyo 133-0033, Japan}
\maketitle
\begin{abstract}
We propose entanglement measures with 
asymptotic weak-monotonicity. We show that a normalized 
form of entanglement measures with the 
asymptotic weak-monotonicity 
are lower (upper) bound for
the entanglement of cost (distillation). 
\end{abstract}
\narrowtext

It is quantum entanglement that led to the
controversy over Einstein-Podolsky-Rosen experiment
\cite{eins} and the non-local nature of quantum mechanics 
\cite{bell}.

The entanglement is the key ingredient in quantum information
processing. The speedup in quantum computation \cite{shor} is obtained
through the parallel quantum operations on massively superposed 
states which are in general entangled.
In particular, in quantum communications such as 
the quantum teleportation \cite{bene}
the entanglement is valuable resource that should not be wasted,
since entanglements can only be obtained by (costly) non-local 
operations.  
For better understanding and manipulation of entangled 
states, we need to classify them as well as 
possible. Quantification of the entanglement-degree 
namely the measure of entanglement is, thus, the central issue in 
quantum information theory \cite{ben2,vede,vida}.

A pair of important measures of entanglement, that is,
the entanglement of cost $E_C$ and entanglement 
of distillation $E_D$ were introduced in a pioneering work 
by Bennett {\it et al.} \cite{ben2}. 
Recently the relation between the
entanglement of cost $E_C$ and  the entanglement
of formation  $E_f$ \cite{ben2} has been clarified \cite{hayd}.
 That is,
$E_C(\rho) = \lim_{n \rightarrow \infty} E_f(\rho^{\otimes n})/n$.
(Here $\rho$ denotes a mixed state.)
The pair of measures of entanglement have been
shown to be the limits for other entanglement measures satisfying 
certain conditions \cite{horo,dona}. 
In other words, any entanglement measure satisfing 
the conditions
is the lower (upper) bound of the entanglement of cost
$E_C$ (entanglement of distillation $E_D$).
One of the conditions for the entanglement measures 
is monotonicity \cite{vida} that
entanglement measures 
cannot be increased by any local quantum operations with
classical communications (LOCCs). 

On the other hand,
it has been shown that there is irreversibility 
in the asymptotic manipulations of entanglement \cite{vid2}.
That is,   
$ E_D(\rho) < E_C(\rho) $ for a class of 
states $\rho$ for which $E_D(\rho)= 0$.
However,
due to possible non-additivity
it is still formidable task to calculate even bounds for 
the $E_C$ and $E_D$ except for a few cases 
\cite{vid2,vid3,shim,hor3,vid4}.
Thus it is not yet clear 
whether $E_C(\rho)$ is strictly greater than 
$E_D(\rho)$ in general.

In this paper, we consider a relaxed and thus conceptually simple
monotonicity requirement, namely asymptotic weak-monotonicity. 
We show that certain re-parameterized 
entanglement measures satisfying the relaxed monotonicity
are the lower (upper) bound for the
entanglement of cost $E_C$ (entanglement of distillation $E_D$). 
These bounds are different from those in Ref. \cite{horo,dona}.
It is notoriously difficult to calculate 
the asymptotic measures of entanglement $E_C$, $E_D$,
or even bounds to them as noted above. 
Thus it is worthwhile for us to give
clues for the bounds to the asymptotic measures of entanglement,
namely the entanglement measures with asymptotic
weak-monotonicity. 

Now let us review basic conditions for the entanglement
measures \cite{hor2}.
(a) Non-negativity: $E(\rho) \geq 0$. 
(b) Vanishing on separable states: $E(\rho)=0$ if $\rho$ is 
separable.
In addition to these, we require monotonicity conditions
\cite{hor2}.
Let LOCCs transform an initial state $\rho_{initial}$ into 
a final state $\rho_{final}$.
The weak-monotonicity is that the entanglement degree does
not increase with LOCCs.
($c.1$) The weak-monotonicity\cite{rudo,horo}:
$E(\rho_{initial}) \geq E(\rho_{final})$.
Let us consider the case where LOCCs
transforms an initial state $\rho_{initial}$ into an 
ensemble $\{p_j, \rho^j_{final}\}$.
Here $j$ is positive integer and
$p_j$ is the probability that $\rho^j_{final}$ will outcome.
The strong-monotoncity says that the expectation value of the 
entanglement degree does not increase with LOCCs.
($c.2$) The strong monotonicity: 
$E(\rho_{initial}) \geq \sum_{j} p^j E(\rho^j_{final})$.
Let us introduce the asymptotic weak-monotonicity.
Consider the asymptotic transformation.
Let us consider transformation of an 
initial state $\rho_{initial}^{\otimes n_i}$ to a final state 
$\rho_{final}^{\otimes n_f}$. ($n$'s are positive integers.)
Here it does not matter whether 
the initial state is transformed to different state 
during the intermediate processings.
Finally, however, the state is reduced to identical copies of
a single state, $\rho_{final}^{\otimes n_f}$.
We say that a state $\rho_{initial}$ can asymtotically be transformed 
to a state $\rho_{final}$, if we can achieve a transformation
$\rho_{initial}^{\otimes n_i} \Rightarrow \rho_{final}^{\otimes n_f} $
at least apporoximately by LOCCs with a condition $n_i \leq n_f$.
More precise definition is the following.

{\bf Definition 1}: We say that a state
$\rho_{initial}$ can asymptotically be transformed to a state 
$\rho_{final}$ with LOCCs, if
the followings are satisfied.
For any $\epsilon, \eta$, and $\delta >0$, there exist an $M$ 
such that for any $n_i \geq M$ 
we can transform a state $\rho_{initial}^{\otimes n_i}$
with LOCCs to a state
$\xi$, with the fidelity
$F(\xi, \rho_{final}^{\otimes n_f(1-\delta)}) \geq 1- \epsilon$ 
and with the success probability $ \mbox{P} \geq 1- \eta$.
Here $n_i \leq n_f$ and the $F(\rho,\rho^{\prime})= 
\mbox{tr} \sqrt{\sqrt{\rho} \rho^{\prime} \sqrt{\rho}} $
is the Uhlmann fidelity \cite{uhlm,jozs}.
$\Box$


{\bf Asymptotic weak-monotonicity condition (c.3)} If an 
initial state $\rho_{initial}$ can be asymptotically 
transformed to a final state
$\rho_{final}$ by LOCCs, then   
$E(\rho_{initial})\geq E(\rho_{final})$. $\Box$

In the asymptotic weak-monotonicity (c.3), exact
transformation is not required: Even though we cannot exactly
transform an initial state to a desired final state, we require that 
$E(\rho_{initial})\geq E(\rho_{final})$
as long as we can asymptotically achieve the transformation.

Let us disscuss on the difference between the entanglement 
measure with asymptotic weak-monotonicity and that of 
strong-monotonicity \cite{vida}.
Consider the case where 
entanglement measures with strong-monotonicity
are transformed by a monotonic function $m$. That is, consider
a re-parameterized function
$E^{\prime}(\rho)= m(E(\rho))$.
Here we assume that the re-parameterized entanglement
measure $E^{\prime}(\rho)$ satisfy
the two obvious conditions (a) and (b).
The re-parameterized function $E^{\prime}(\rho)$
is no longer the entanglement measure with strong-monotonicity
in general:
the entanglement measure involves with processes
where the number of distinguishable states varies. For example,
measurements can transform
an initial state into different states.
In mixing processes, different states are transformed to 
a single state. In the both cases, strong-monotonicity requires
that the expectation value of entanglement
cannot increase. However, it is 
easy to see that the re-parameterization does not preserve
strong-monotonicity in general with respect to the processes
where the number of the states varies.  
However, it is clear that the asymptotic
 weak-monotonicity is preserved
upon the re-parameterization, since we are considering the 
one-dimensional orderings in this case. 

{\bf Proposition 1}: 
The asymptotic weak-monotonicity is preserved upon
the re-parameterization. Namely, if $E(\rho)$ is an entanglement
measure with asymptotic weak-monotonicity and $E^{\prime}(\rho)= m(E(\rho))$
where $m$ is a monotonic function, then $E^{\prime}(\rho)$ is
also an entanglement measure with asymptotic
weak-monotonicity. $\Box$


{\bf Proposition 2}: 
All entanglement measures with asymptotic 
weak-monotonicity gives rise to the 
same orderings for pure states. 
\\
{\em Proof}: 
Let us assume two entanglement measures $E_i$ ($i= A, B$)
with asymptotic
weak-monotonicity give rise to different orderings for 
two pure states $\rho$ and 
$\rho^{\prime}$.
The fact that the
order is reversed in dependence on entanglement measures 
obviously means that
entanglement-degree of $\rho$ is less than that of $\rho^{\prime}$
in one of the two measures and vice versa in the other.
That is, we have either
\begin{eqnarray}
\label{a}
E_A(\rho) > E_A(\rho^{\prime})
\hspace{2mm} \mbox{and} \hspace{2mm}
E_B(\rho) < E_B(\rho^{\prime}),
\end{eqnarray}
or
\begin{eqnarray}
\label{b}
E_A(\rho) < E_A(\rho^{\prime})
\hspace{2mm} \mbox{and} \hspace{2mm}
E_B(\rho) > E_B(\rho^{\prime}).
\end{eqnarray}
Due to the asymptotic weak-monotonicity condition (c.3) and
Eqs. (1) and (2), the state 
$\rho$ can neither be asymptotically
transformed to the state $\rho^{\prime}$
nor $\rho^{\prime}$ to $\rho$.
That is,
the two states $\rho$ and $\rho^{\prime}$ are 
(asymptotically) incomparable \cite{niel,ben3,hwan}.
However, for the two pure states $\rho$ and $\rho^{\prime}$,
we can achieve the asymptotic transformation of either
$\rho^{\otimes n} \Rightarrow (\rho^{\prime})^{\otimes n} $ or 
$\rho^{\otimes n} \Leftarrow (\rho^{\prime})^{\otimes n} $,
by entanglement concentration and dilution \cite{ben4}.
This is in contradiction with the former statement.
$\Box$

The Proposition 1 says that we can generate numerous 
entanglement measures with asymptotic weak-monotonicity from
a single one.
On the other hand, the unique 
entanglement measure for pure states is also an entanglement
measure with asymptotic weak-monotonicity 
for pure states. 
Here the unique measure for a pure state is the followings.  
$E_p(|\Psi\rangle \langle \Psi|) 
= S(\mbox{Tr}_B |\Psi\rangle \langle \Psi| )$,
where $S(\rho)= \mbox{Tr}(-\rho \log_2 \rho)$ and $B$ 
denotes the 
latter one of the two parties Alice and Bob
\cite{pope,vida}.
Thus by the proposition 2 for pure states
the ordering of the entanglement measures with asymptotic
weak-monotonicity is the same as that of the unique 
measure of entanglement.
Therefore, we can fix the freedom of entanglement measure $E$ 
with asymptotic weak-monotonicity involved with the 
Proposition 1 by the following condition.

{\bf Normalization condition}: For any pure state $\rho$,
$E(\rho)$ is 
the same as the unique entanglement measure $E_p(\rho)$
for pure state \cite{pope,vida}. 

Although the normalization condition
is involved with entanglement-degrees only for pure states,
it is also fixing those for mixed states since entanglement-degrees
of pure states 
are continously distributed \cite{virm}. 
We denote the entanglement measure with
asymptotic weak-monotonicity
satisfying the normalization condition as $\tilde{E}$.

{\bf Proposition 3}: The normalized entanglement measure 
with asymptotic weak-monotonicity
$\tilde{E}(\rho)$ is the lower bound for 
the entanglement of cost 
$E_C(\rho)$. \\
{\em Proof}: 
Assume that there exists a state $\rho$ such that 
$ E_C (\rho) < a < \tilde{E}(\rho)$. 
Then we choose
a pure state such that
$E_p(|\Psi\rangle \langle \Psi|)= a$.
This means that we can asymptotically distill
$m a$ numbers of the 
maximally entangled states 
$(|\phi^{+}\rangle \langle\phi^{+}|)^{\otimes ma}$
from the $m$ copies of the state 
$|\Psi\rangle \langle \Psi|$.
(Here
$|\phi^+\rangle= (1/\sqrt{2})(|00\rangle +|11\rangle)$.)
Then some of the $m a$ numbers of 
the maximally entangled states
can be asymptotically transformed to 
$\rho^{\otimes m}$ \cite{ben3},
since $m {E}_C (\rho) < m a$.
What we have done is the asymptotic transformation
$|\Psi\rangle \langle \Psi| \Rightarrow \rho$.
By the definition 1, we have
$\tilde{E}(|\Psi\rangle \langle \Psi|) \geq \tilde{E}(\rho)$.
(Note that the continuity condition is not additionally necessary
here while it is in the case of Ref. \cite{horo,dona}.
The asymptotic weak-monotonicity
is defined such that the continuity condition is not necessary.)
By the normalization condition, we have 
$\tilde{E}(|\Psi\rangle \langle \Psi|)= a$.
Thus we have 
$\tilde{E}(|\Psi\rangle \langle \Psi|) < \tilde{E}(\rho)$,
which contradicts above obtained inequality.
$\Box$

We can get a proposition for 
the entanglement of distillation 
$E_D(\rho)$ in a similar way. 

{\bf Proposition 4}: The normalized entanglement measure 
with asymptotic weak-monotonicity
$\tilde{E}(\rho)$ is the upper bound for 
the entanglement of distillation 
$E_D(\rho)$. \\
{\em Proof}: 
Assume that there exists a state $\rho$ such that 
$\tilde{E}(\rho) < a <  {E}_D (\rho) $.
However, we can asymptotically distill
$m E_D (\rho) $ numbers of the 
maximally entangled states from the state
$\rho^{\otimes m}$, by definition. 
Then we choose a pure state such that
$E_p(|\Psi\rangle \langle \Psi|)=a$.
Some of the $m E_D (\rho)$ numbers of 
the maximally entangled states
can be asymptotically transformed to 
$(|\Psi\rangle \langle \Psi|)^{\otimes m}$,
since $m E_D (\rho) > m a$.
Thus, we have done asymptotic transformation
$\rho \Rightarrow |\Psi\rangle \langle \Psi|$.
By defintion 1, we have
$\tilde{E}(\rho) \geq \tilde{E}(|\Psi\rangle \langle \Psi|) $.
By the normalization condition, we have 
$\tilde{E}(|\Psi\rangle \langle \Psi|)=a$.
Thus we have 
$ \tilde{E}(\rho) < \tilde{E}(|\Psi\rangle \langle \Psi|) $,
which contradicts the above obtained inequality. 
$\Box$
 
The entanglement measures with strong-monotonicity
satisfying certain additional conditions in Ref. \cite{horo,dona}
are bounds for the asymptotic measures, namely the  
entanglement of cost $E_C$ and entanglement of distillation $E_D$. 
As we have shown, the entanglement measures with
asymptotic weak-monotonicity
satisfying an additional normalization condition
are also bounds for the asymptotic measures. 
Due to the existence of the additional conditions,
the two types of bounds can be different.

In conclusion, we have proposed entanglement measures with
asymptotic weak-monotonicity.
The orderings of the entanglement measures $E(\rho)$ with 
asymptotic weak-monotonicity for the pure states
is the same as that of the unique measure
of entanglement $E_p(\rho)$ for the pure states \cite{pope,vida}. 
This fact enabled us to re-parameterize entanglement
measures with asymptotic weak-monotonicity $E(\rho)$
such that the $E(\rho)$ coincides with $E_p(\rho)$
for all pure states.  
We have shown that normalized entanglement measures
with asymptotic weak-monotonicity $\tilde{E}$
are lower (upper) bound for 
the entanglement of cost $E_C$
(entanglement of distillation $E_D$). 

\acknowledgments
We are very grateful to Prof. Hiroshi Imai and
Japan Science Technology Cooperation
for financial supports. We are also very grateful to Dr. G. Vidal
for helpful discussions.

\end{document}